\documentclass[conference]{IEEEtran}
\pdfoutput=1

\ifCLASSINFOpdf
\else

\fi

\usepackage{booktabs}
\usepackage{enumitem}
\newcommand{\changeitem}{%
  \let\latexitem\item
  \renewcommand\item[1][]{\latexitem\relax{##1 --} }%
}

\newlist{steps}{enumerate}{1}
\setlist[steps]{label=\textit{Step \arabic*:},leftmargin=*}

\newenvironment{labelledsteps}[1][]
  {\begin{steps}[before=\changeitem,#1]}
  {\end{steps}}
\usepackage{varwidth}
\usepackage{capt-of}
\usepackage{listings}
\usepackage{amsmath}
\usepackage{framed}
\usepackage{tabu}
\usepackage{float}
\usepackage[caption = false]{subfig}

\usepackage{color}
\usepackage{graphicx}
\usepackage{cleveref}
\usepackage{fancybox}
\usepackage[export]{adjustbox}
\usepackage[table]{xcolor}
\definecolor{Gray}{gray}{0.9}
\definecolor{White}{RGB}{255,255,255}
\usepackage{bbold}
\usepackage{bbm}
\usepackage{mathbbol}
\usepackage{multirow}
\usepackage{fix-cm}
\usepackage{siunitx,booktabs}
\usepackage{xcolor}
\usepackage{csquotes}
\usepackage{setspace}
\usepackage{graphicx}
\usepackage{colortbl}
\usepackage{array}
\usepackage{pifont}
\usepackage {rotating}
\usepackage{mwe}
\usepackage{pbox}
\usepackage{enumitem}
\let\oldding\ding
\renewcommand{\ding}[2][1]{\scalebox{#1}{\oldding{#2}}}

\newcommand{\Csharp}{%
  {\settoheight{\dimen0}{C}C\kern-.05em \resizebox{!}{\dimen0}{\raisebox{\depth}{\#}}}}

 \usepackage{enumitem}

\usepackage{xspace}
\usepackage{xparse}
\DeclareDocumentCommand\newstep{o}{%
\item\IfNoValueTF{#1}{}{#1 \textendash\xspace}}

\definecolor{orange}{RGB}{0,32,96}
\definecolor{o}{RGB}{245,245,245}
\definecolor{g}{RGB}{50,50,50}

\usepackage{cite}
\usepackage{capt-of}

\PassOptionsToPackage{gray}{xcolor}
\usepackage{tikz}

\usepackage{tcolorbox}

\begin{document}
%
\title{\LARGE Learn More, Pay Less! Lessons Learned from Applying the Wizard-of-Oz Technique for Exploring Mobile App Requirements}

\author{\IEEEauthorblockN{Zahra Shakeri Hossein Abad, Shane D.V. Sims, Abdullah Cheema, Montasir B. Nasir, Payal Harisinghani}
\IEEEauthorblockA{Department of Computer Science\\ University of Calgary, Calgary, Canada\\
Emails: \{zshakeri, shane.sims, cheemaab, mbechirn, payal.harisinghani\}@ucalgary.ca}}


\maketitle


\begin{abstract}
Mobile apps have exploded in popularity, encouraging developers to provide content to the massive user base of the main app stores. Although there exist automated techniques that can classify user comments into various topics with high levels of precision, recent studies have shown that the top apps in the app stores do not have customer ratings that directly correlate with the app's success. This implies that no single requirements elicitation technique can cover the full depth required to produce a successful product and that applying alternative requirements gathering techniques can lead to success when these two are combined. Since user involvement has been found to be the most impactful contribution to project success, in this paper we will explore how the Wizard-of-Oz (WOz) technique and user reviews available in Google Play, can be integrated to produce a product that meets the demand of more stakeholders than either method alone. To compare the role of early interactive requirements specification and app reviews, we conducted two studies: (i) a case study analysis on 13 mobile app development teams who used very early stages Requirements Engineering (RE) by applying WOz, and (ii) a study analyzing 40 (70, 592 reviews) similar mobile apps on Google Play. The results of both studies show that while each of WOz and app review analysis techniques can be applied to capture specific types of requirements, an integrated process including both methods would eliminate the communication gap between users and developers at early stages of the development process and mitigates the risk of requirements change in later stages.

%
%


\end{abstract}



\IEEEpeerreviewmaketitle




\section{Introduction}

In many respects, developing mobile applications (apps) is similar to software engineering for other embedded applications \cite{Eng}. However, apps often have requirements that are not associated with traditional software such as the high-level user interaction and the need for an intelligent and intuitive User Interface (UI) design. These requirements pose a great challenge to current app developers and have forced software engineers to re-evaluate current engineering practices when dealing with mobile apps.
Research and industry documentation provide evidence that traditional methods of Requirements Engineering (RE) practice are ineffective. A recent study by Erfani J. et al. \cite{Mesbah} reveals that requirements for mobile app projects change rapidly and often over long periods of time, contributing to changes in the UI and logic of app operations. 

While there are well-known techniques for managing requirements change in various situations (e.g. as in \cite{APSEC, SERA, ernst2012case}), the high level of requirements change and the high number of features on mobile apps makes the process of eliciting, evaluating, and validating mobile app requirements complicated and error prone. Some problems with just-in-time RE is that it provides no explicit big picture of the requirements, prioritization phase, requirements elicitation, and unclear feature provenance \cite{ernst2012case}. This is in spite of the fact that requirements elicitation has been found to be one of the most critical processes in RE. In the elicitation process, involving the right stakeholders has been found to be crucial. This is especially true of user-end products like mobile apps. Failing to involve the right stakeholders, such as users, can result in incomplete and incorrect requirements \cite{razali2011selecting}. The most important aspect of a requirements elicitation technique is its ability to transfer knowledge from users to analysts \cite{anwar2012practical, IWSPM}. Developers should allow users to take part in the elicitation process in the original design and subsequent iterations of the project as studies have shown that 44\% of users will immediately delete applications with poor performance \cite{inukollu2014factors}. The Wizard of Oz (WOz) technique is a low-fidelity prototyping technique where software requirements are simulated to give an impression of how these requirements will work when actually implemented. The simulation is usually by way of paper sketches of the interface that are moved and changed by a facilitator in response to the user's input. The WOz technique is of interest to us because it can elicit requirements that are not possible to be gathered by other requirements elicitation techniques. It allows users to be involved in the requirements process in a way that is not possible by other elicitation techniques \cite{wik2015using}. The WOz method can facilitate designers who experience blocks in their ability to come up with novel ideas and avoids inaccurate user preference assumptions, as WOz can be done prior to building complete prototypes \cite{dow2005wizard}. Resultantly, we will explore how WOz can be used to produce better applications. To compare the role of user review analysis and WOz in eliciting and defining mobile app requirements, we defined the following main Research Questions (RQs):
\vspace{-1mm}
\begin{itemize}
\item {\bf RQ1: How does the WOz technique help in capturing mobile app requirements?}\\
This RQ aims to explore the role of the WOz technique in capturing mobile app requirements during the early stages of app development. Moreover, this RQ investigates the contexts best served by the WOz approach.

\item{\bf RQ2: How do WOz and user review analysis techniques differ in identifying mobile app requirements?}\\
This RQ compares the requirements captured by applying each of the WOz prototyping and user review analysis techniques. The results of this RQ will help requirements engineers and app developers gain more insight into the applicability of each technique for exploring the requirements of mobile apps in various stages of the development process. 
\end{itemize}

To address these RQs, we gather data on the efficacy of applying of each of the above-mentioned techniques to capturing mobile app requirements. This is done by conducting two studies: (1) a field study on 13 Android app development teams (Section \ref{sec:E1}), and (2) a retrospective data analysis on user reviews (Section \ref{sec:E2}). Our analysis of the WOz technique found that it is highly effective in capturing requirements that were missed in earlier stages of requirement collection and is most effective in capturing Non-Functional Requirements (NFRs), with all teams in the study having made at least one change to their app's NFRs. The retrospective data analysis we conducted involved analyzing the reviews of 40 mobile apps (similar to the apps studied in Experiment 1) in the Google Play Store. From this study we gained insights into the differences between the two elicitation techniques. We conclude both studies with a discussion of situations in which each technique might help developers explore mobile app requirements (Sections \ref{sec:LL1} and \ref{sec:LL2}). This paper makes the following main contributions:

\begin{itemize}[leftmargin=2.3\labelsep]
\item It presents results from a field study of 13 app development teams who applied the WOz technique for capturing app requirements in early stages of their development process.
\item It presents results of a retrospective analysis of user reviews (\(70,592\) reviews) of 40 apps available on Google Play.
\item It compares the application of WOz and app review analysis for capturing mobile app requirements. 
\end{itemize}


\section{Related Work}
\label{sec:RW}

This section presents related research on capturing and managing requirements of mobile apps. In particular, we report on research works addressing RE techniques for exploring mobile app requirements, mobile app usability, and the application of prototyping and WOz in RE activities.


To explore the main topics of user reviews on mobile app stores, Carre{\~n}o and Winbladh \cite{RE1} applied the Topic Modelling technique and evaluated the validity of using this technique in the context of requirements evolution. To this end, they applied the Aspect and Sentiment Unified Model (ASUM) approach, which is an extension of Latent Dirichelt Allocation (LDA) and includes sentiment classification. The results of this study show that the application of this approach clearly presents the main topics relevant to requirements changes. Likewise, Chen et al \cite{RE5} proposed a computational framework called ``AR-Miner'' to facilitate exploring and visualizing informative user reviews, classifying and prioritizing these reviews by applying the Topic Modelling approach, and a review ranking scheme. This framework provides various types of informative (e.g. functional and performance flaws, and requirements change) and non-informative  (e.g. pure user emotional expressions, app descriptions, and questions and inquiries) information for app developers and requirements engineers. 
In a similar study, Pagano and Maalej \cite{RE3} conducted an empirical study to understand the role of app stores as a forum for communication between users and developers, and to explore the methods and tools for analyzing and aggregating user reviews. They applied descriptive statistics and frequent item-set mining for analyzing the usage of user reviews and identifying latent patterns among topics, respectively. The results of this study imply the important role of app stores as a communication channel among users and app developers and as a rich repository for understanding user need, bugs, user experience, and feature requests.

In a recent study, Buchmann and Karagiannis \cite{RE2} proposed a modelling method which enables semantic traceability for representing requirements and supports the requirements elicitation of mobile apps. This method provides a knowledge externalization channel between app developers and business stakeholders and bridges the twin peaks of requirements collection and early designs. The authors employed a meta-modelling framework to represent process-centric app requirements, which can be semantically traced based on early design aspects or context-dependent elements. This study shows that requirements representation at early stages of the development process should receive as much emphasis as other aspects of requirements management.

In regards to mobile app usability, Nayebi et al. \cite{RE6}, surveyed the literature on usability features to evaluate mobile app usability and determined that there is inadequate scientific research addressing the requirements of a new mobile user interface. They also found that there is a clear need for a field study methodology analyzing users behaviour during their interaction with mobile apps. Baharuddin et al. \cite{RE7} proposed twenty-five usability dimensions as a guideline for designing and evaluating mobile apps. They prioritized these dimensions based on the results of 9 empirical studies. Effectiveness, efficiency, satisfaction, usefulness, and aesthetic are the top five dimensions on this list.  

Regarding the application of the WOz technique in RE activities, Molin \cite{OZ1} conducted a study evaluating WOz effectiveness in specifying requirements of interactivity in multimedia products. The results of this study show that WOz prototyping provides a true interactive experience and supports and extends the requirements specification. Li et al. \cite{OZ2} developed a WOz tool for testing location-enhanced applications. This tool provides a set of high-level prototyping abstractions (e.g. maps, and storyboards), which can be used to model the location of people, places, and things. This helps developers of location-based applications to rapidly explore various possible designs in the early stages of the development. However, the authors believe that because the interaction logs and user behaviour data generated by this tool is in a large scale, the data still needs to be analyzed and visualized.

While the existing research provided a wealth of insight on the application of data analysis, crowdsourcing, and prototyping for exploring the requirements of mobile apps, we could not find any study that investigated the differences between the application of WOz compared to user reviews in this context. Moreover, the comprehensiveness of our study in terms of conducting two separate studies to understand both methods makes it different from other investigations.

\begin{figure}
\centering
\subfloat[UofC Event Planning App]{\includegraphics[scale=.2]{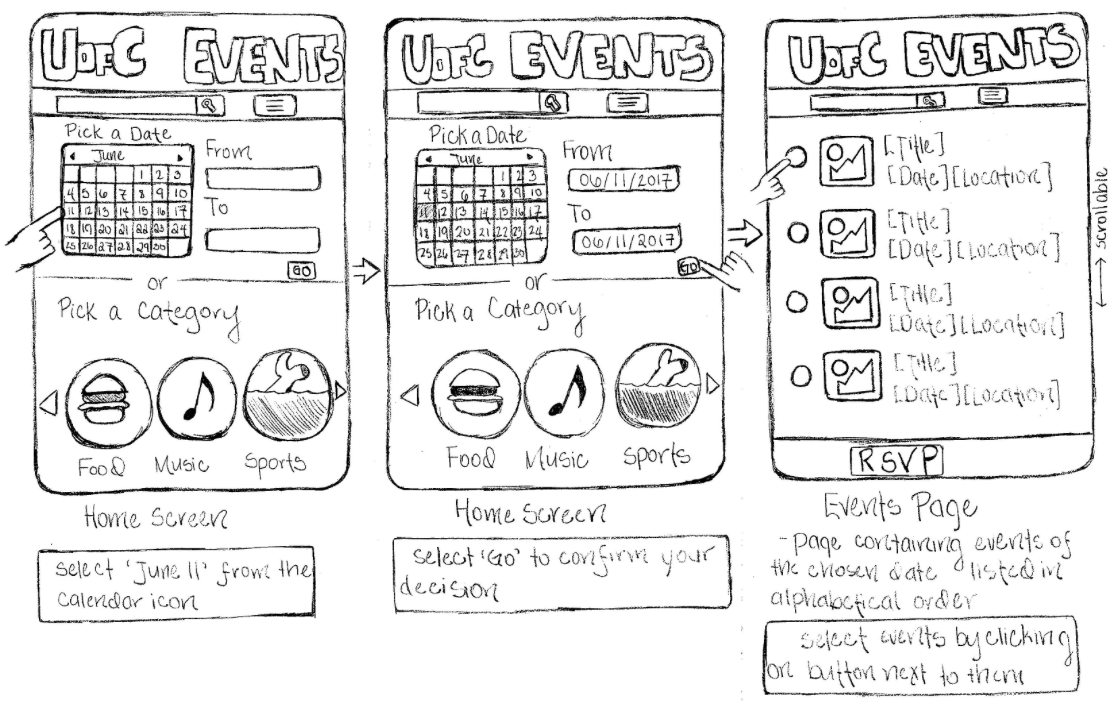}}\hspace{1em}\\ [-.05em]
\subfloat[Event Parser App]{\includegraphics[scale=.15]{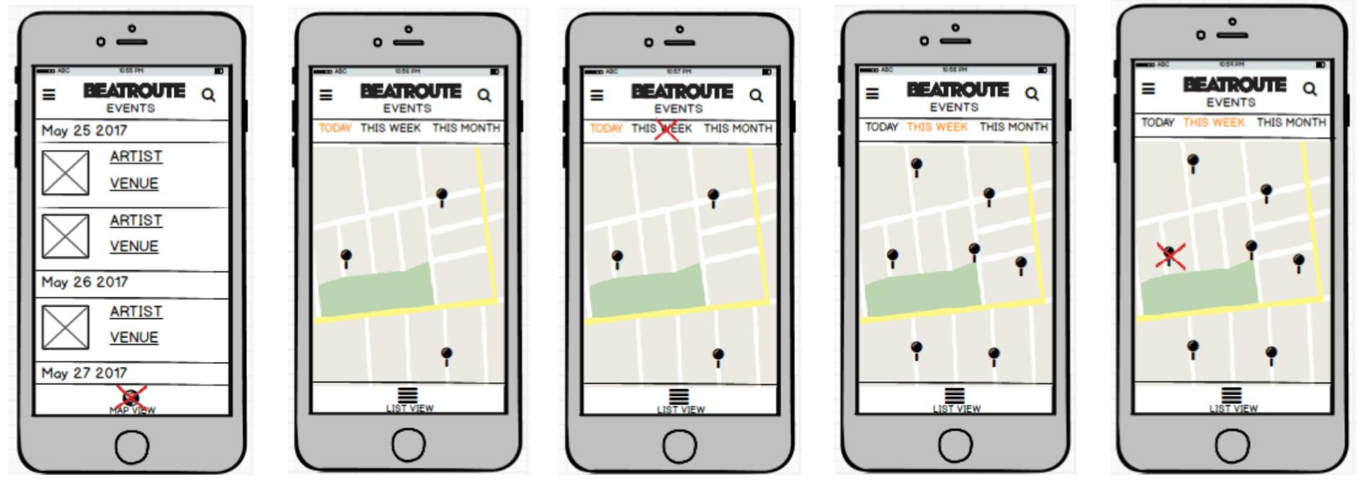}}
\caption{Sample storyboards generated by the development teams under study}
\label{fig:SB}
\vspace{-5mm}
\end{figure}

\section{Experiment 1- The Role of Wizard-of-OZ in Requirements Elicitation Activity}
\label{sec:E1}
In this section we explore the efficacy of WOz in capturing requirements in early stages of the software development process. Specifically we look at the role that WOz played in capturing requirements for mobile app design. As storyboarding (Figure \ref{fig:SB} a-b) was done as a precursor to (and a basis for) WOz, we have grouped these low-fidelity prototyping methods together in the following discussion. This study has been approved by the University of Calgary Conjoint Faculties Research Ethics Board (CFREB6\footnote{
http://www.ucalgary.ca/research/researchers/ethics-compliance/cfreb}).
\subsection{Data Collection and Preparation}
\subsubsection{Data Collection}
Data for this experiment was collected during a three week study from fifteen mobile app development teams in the initial stages of the development of an Android application. Each team had two or three members, with the number of participants totalling forty-five. The results of two of the teams were omitted from the study for lack of applicability. Further, the results surrounding new Functional Requirements (FR) from one team were omitted as these results were not, the opinion of the researchers, typical or representative; their inclusion causing the resulting data to be misleading (twelve new FRs for the omitted team vs. one for the team with the next highest new FR).

\begin{figure*}
\vspace{-5mm}
\centering
\subfloat[Flash Card App]{\includegraphics[scale=.19]{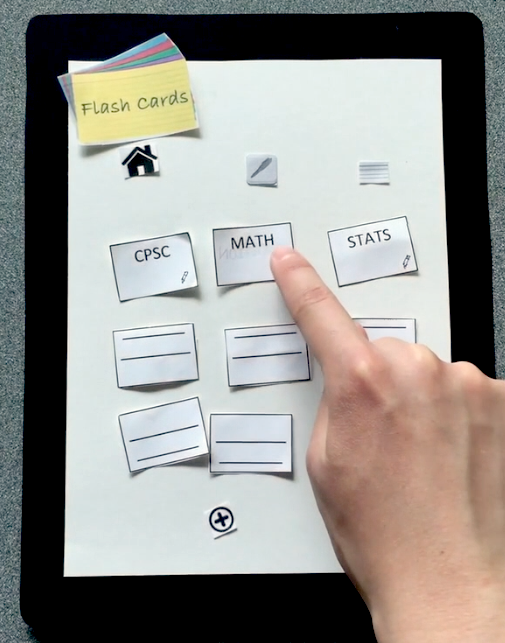}}\hfill
\subfloat[Express Food Ordering App]{\includegraphics[scale=.18]{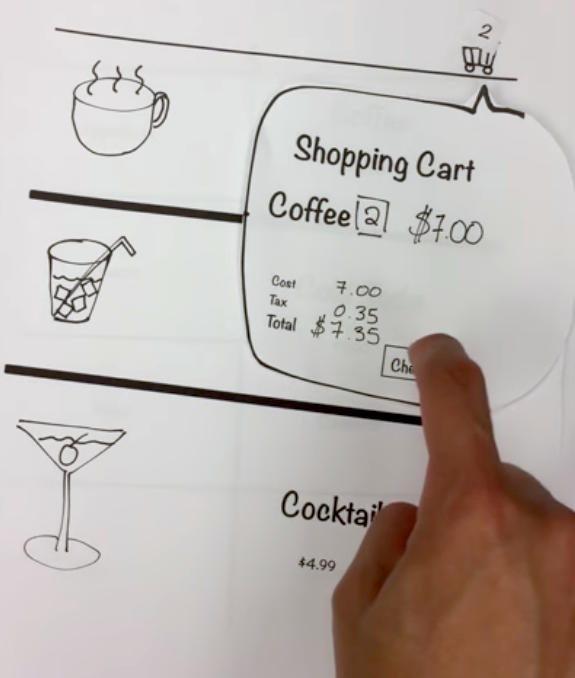}}\hfill
\subfloat[UofC Fitness App]{\includegraphics[scale=.165]{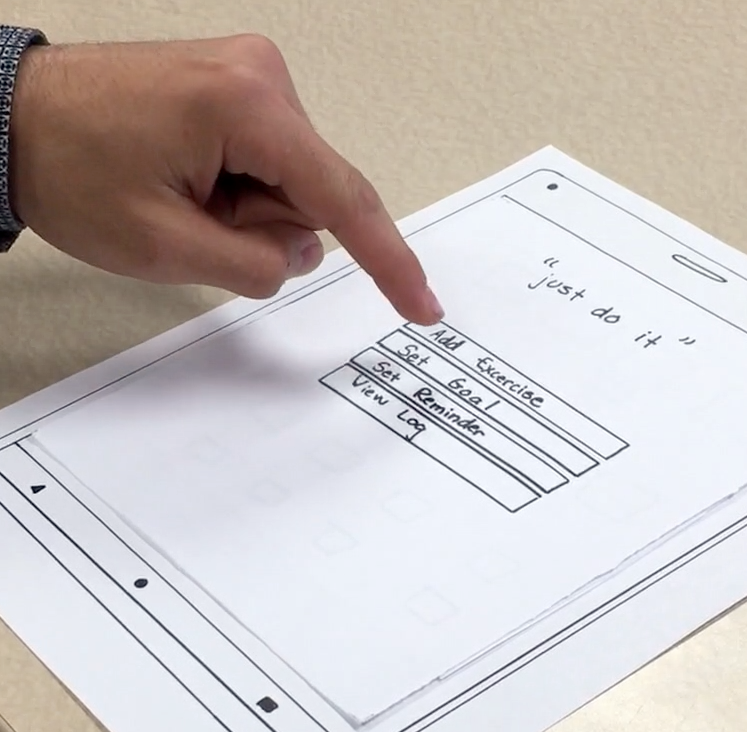}}\hfill
\subfloat[Data Collection App]{\includegraphics[scale=.19]{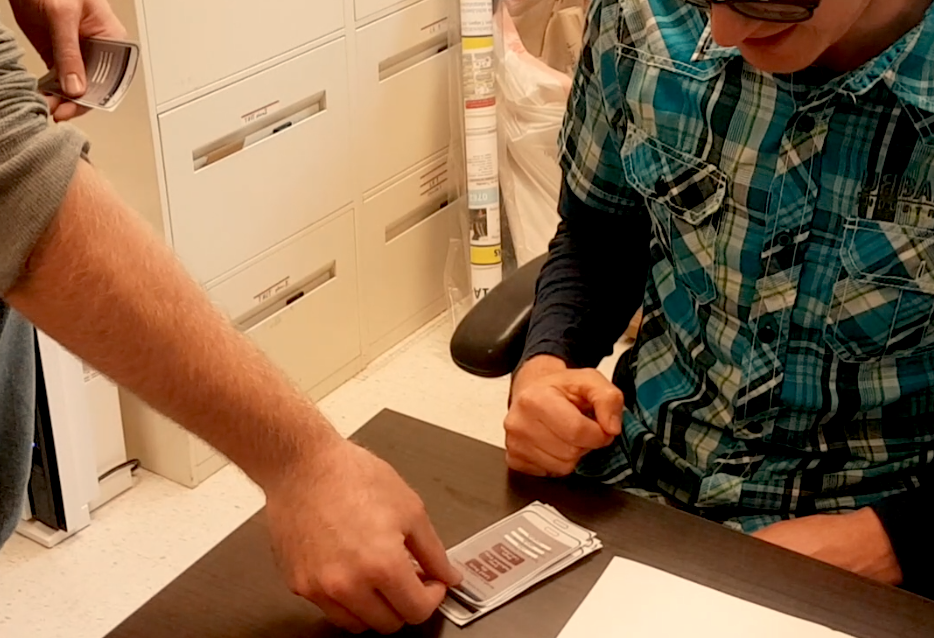}}\hfil
\caption{Screenshots of the process of implementing WOz for exploring/revising the requirements of apps under study}
\label{fig:WOz}
\vspace{-5mm}
\end{figure*}

 Teams were provided with one week of instruction in RE processes and techniques, including low-fidelity prototyping. Over the course of the following two weeks, teams applied what they learned in the instruction phase to collecting requirements for a new mobile app. Requirements were initially collected from client product descriptions and meetings between the development teams and their respective clients. Following this initial collection phase (which serves as a baseline for comparison with WOz), teams employed WOz in an attempt to ensure that all requirements had been collected from the client and understood by the team. WOz was conducted by interaction between the development team and client on one hand and the low-fidelity application on the other. Additions and changes to requirements were then recorded and submitted as responses to the research team. Figure \ref{fig:WOz} (a-d) represents the implementation of the WOz technique by four of the development teams under study.

\subsection{Data Analysis}
\label{sec:RQ1-Analysis}
Data was analyzed by applying basic statistical methods and a revised version of the grounded theory methodology \cite{GT}; the later used to extract meaningful data from the thirteen development team responses included in this study. Grounded theory methodology requires us to analyze each development team response line-by-line, to extract meaningful results. In this context meaningful results were those statements concerned (explicitly or implicitly) with FRs, NFRs, and statements about WOz as it pertained to the development team in question. To aid us in this process, the web-based coding tool Saturate \footnote{http://www.saturateapp.com/} was used. Saturate allowed us to trace between codes and the applicable data. The process used to analyze each of the team responses was as follows:

Each response was numbered by team and their respective apps were coded to represent one of nine descriptive categories (see Table \ref{tab:AppData}) based on the primary usage of the app. This represents the top level of our coding hierarchy. Each response was then selected in turn and read to generate a set of high level categories in a process known as open coding. This identified five distinct themes: an instance of a new FR, revised FR, new NFR, revised NFR, or other data point of interest; the latter used to capture statements of significance concerning WOz. For each data-point in the text (within the scope of this study) a coding was applied. Completing this for each response, while maintaining the option of going back to previously read responses to enforce coherence of concepts and categories, allowed us to achieve theoretical saturation. Theoretical saturation describes the state that results when no new concepts, information or relationships are revealed by continued use of the methods just described \cite{GT2}. The results of applying this modified grounded theory technique are presented in Section \ref{sec:RQ1Results}.

\subsection {Results}
\label{sec:RQ1Results}

\begin{figure*}
\begin{framed}
\vspace{-3mm}
\centering
\vspace{-3mm}
\subfloat[Requirements captured (new vs refined)]{\includegraphics[scale=.55]{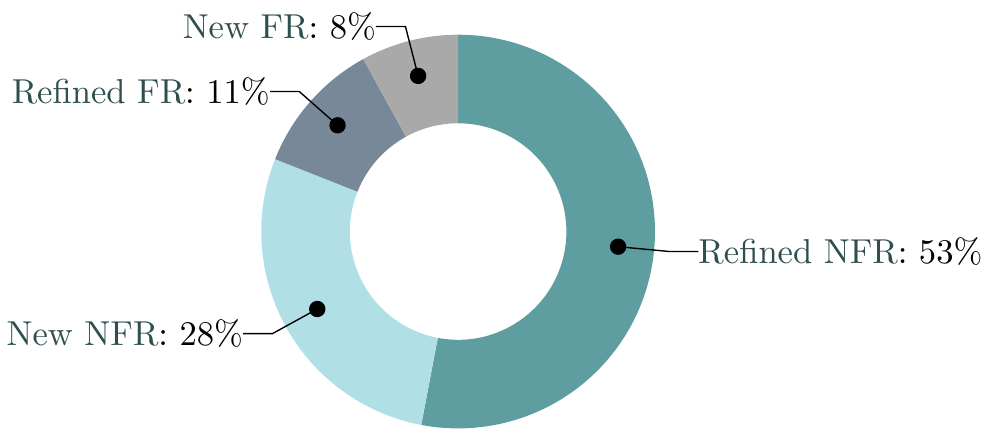}}\hfill
\subfloat[Captured NFRs]{\includegraphics[scale=.53]{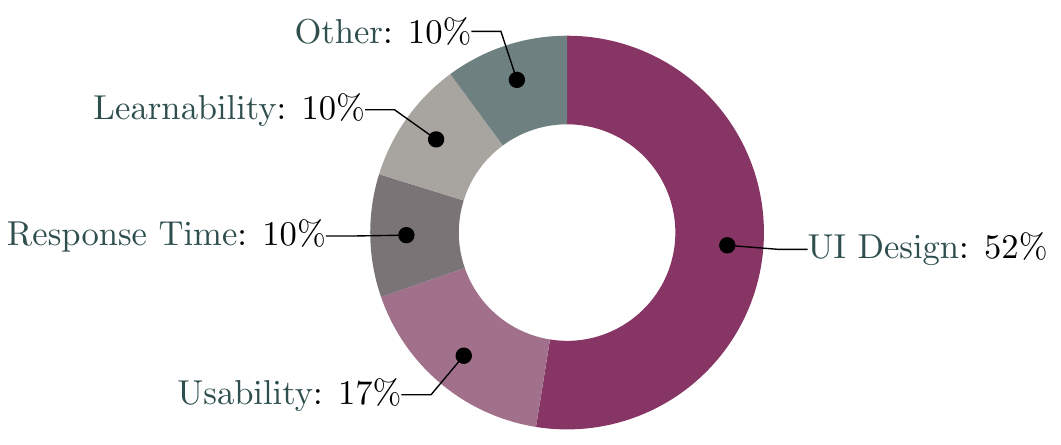}}\hfill
\subfloat[Requirement collection by App type]{\includegraphics[scale=.53]{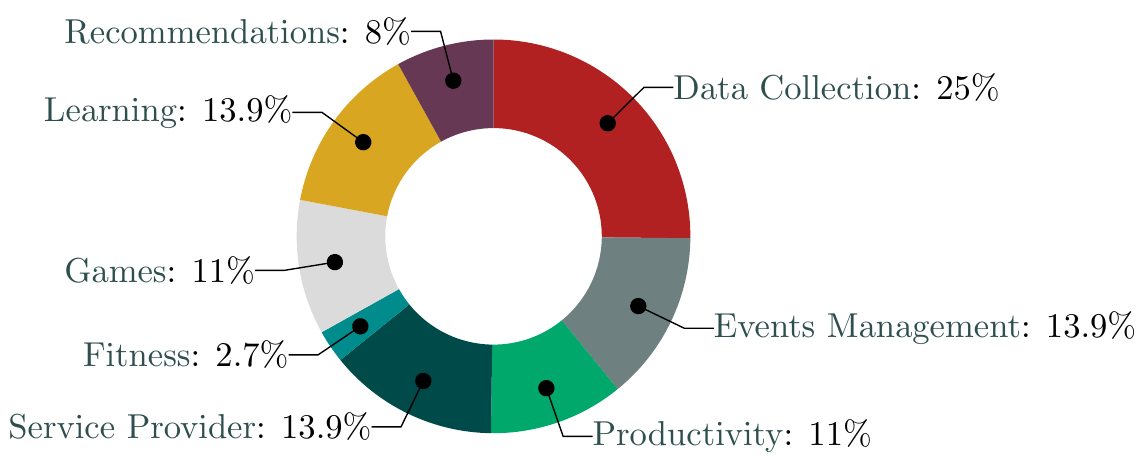}}\hfill
\vspace{-3mm}
\end{framed}
\vspace{-3.5mm}
\caption{RQ1- application of the WOz technique for capturing mobile app requirements}
\label{fig:RQ1}
\vspace{-5mm}
\end{figure*}

Of the teams surveyed, all reported changes to captured NFRs while 46\% of the {\it teams} also made changes to their FRs. As illustrated in Figure \ref{fig:RQ1}a, of the requirement {\it changes}, 81\% were NFRs (53\% of total changes were refinements to existing NFRs, 28\% were newly captured NFRs). Only 19\% of changes were to FRs (8\% new FRs and 11\% refinements to existing FRs).  Of the changes to the NFRs, approx. 53\% were changes related to UI design (all but one team made a change to UI design), 17\% where changes related to usability, 10\% of changes were related to response time, 10\% were changes related to learnability, while 10\% of the changes fall into other categories (Figure \ref{fig:RQ1}b).

Regarding the types of apps that were best served by the WOz approach for requirements capturing, as illustrated in Figure \ref{fig:RQ1}c, of the requirements captured following the WOz technique, the changes by app type were as follows: 25\% by Data Collection apps, 14\% by Events apps, 14\% by Service apps, 11\% by Game apps, 13.9\% by Learning apps, 2.7\% by Fitness apps, 8\% by Recommendation apps and 11\% by Productivity apps. Given the small sample size, no conclusions can be drawn from the results above, so far as determining which (app type) context the WOz works best for. However, the results do suggest one interesting insight. It may be that WOz works best in collecting requirements in the following situations:

\begin{itemize}
\item Clients have themselves been unclear/unsure of the requirements; or,
\item The app to be developed is of a type that sees infrequent use by the general population (developers included) and is thus less likely to have intuitive requirements.
\end{itemize}
\vspace{-1mm}
\subsection{Lessons Learned and Challenges}
\label{sec:LL1}
 In general it can be said that WOz was highly effective in capturing requirements missed in earlier stages of collection. In fact all of the teams captured or clarified at least one requirement after completing WOz prototyping. Further, the results show that WOz was most effective in capturing NFRs, with all teams having made at least one change to their app NFRs. A possible explanation of this is that clients are more likely to have thought about the functions their application should be capable of (their motivation for soliciting its creation), as opposed to ways to judge how the system should operate.

A point of interest is that when a NFR was captured following WOz, it was much more likely to result in a refinement to an existing NFR rather than a new NFR (66\% vs. 34\% respectively). Meanwhile, when a FR was captured, this disparity was less pronounced (43\% new FRs vs 57\% refinements). This intuition is echoed in the following statement captured from a development team response:
 { \it ``We began our UI Planning with very clear requirements communicated to us [É]. We relied on those requirements as guidelines for the visual design of the application. Yes [WOz] caused us to revise certain elements of the design, but those revisions were to the design alone - not the actual requirements the design was meant to service''}.
 
 In fact this team did not have any changes to their FR following WOz. This statements suggests that the more clear the client's ideas are as to how their app will function and the more clearly this is communicated with the development team, the less likely missed (functional) requirements are to be captured with WOz. But as the statements notes, despite these ideal circumstances surrounding clarity of requirements by the client, WOz was nonetheless able to capture revisions to UI design elements. The following statement from another development team expresses a similar sentiment:
 
 {\it ``Overall, [WOz] didn't heavily influence core functional/non-functional requirements \textelp{}, but it did help us realize what UI design decisions we did and didn't like, which we will continue update until we have a UI that is aesthetically pleasing and functional.''}

Moreover, it was noted that the WOz technique resulted in changes to NFRs by 100\% of teams. Of these captured NFRs, deficient or missing NFRs involving UI design were captured by WOz prototyping more than any other single category of NFRs at 52\% of NFRs captured. The next largest category accounted for 14\% of captured NFRs. If UI design, learnability and usability are grouped together and thought of as design requirements, then a large majority (80\%) of NFRs captured by WOz fall into this category. As one might expect, this suggests that WOz is most effective in capturing requirements that involve anything to do with the user experience. This finding is in keeping with the pronounced visual and simulated interactive nature of the WOz prototyping technique.

\subsection{Threats to Validity}

In answering RQ1 concerning the efficacy of the WOz technique, the methodologies of qualitative analysis as provided for in grounded theory were adhered to with as much rigour as the circumstances of the study would allow. That being said, there are a number of threats to the validity of the results presented above. All of these threats originate in the \textit{``Data Collection and Preparation''} phase of this study. 

The first of these threats is the context in which the development teams were working. Each development team was composed of Software Engineering students and thus not industry practitioners in most cases. This relates to another concern, namely that this study was the first experience with RE that most of the participants are likely to have had. The concern with this is the potential that perhaps a significant number of the requirements missed in the pre-WOz stages of requirements gathering may not be indicative of what would be experienced by industry development teams.

Finally, the small sample size represents a potential threat. As noted, this study was conducted with forty-five participants broken down into fifteen teams, resulting in responses being collected on thirteen apps. While the number of participants in this study is in keeping with the parameters generally accepted in qualitative analysis \cite{patton1990qualitative}, it is not \textit{clear} that these numbers will have resulted in participant saturation; the point at which the inclusion of additional participants increases the size of the data but produces no significant changes in the results. The sample size also caused some disparity in the number of apps per category, resulting in more potential data points for some categories than others, though coverage was reasonable.

\section{Experiment 2 - WOz vs. User Reviews}
\label{sec:E2}
\subsection{Data Collection and Preparation}
From the product descriptions provided by the clients, apps were matched by FR and NFR to apps on Google Play. In addition to matching similarities, apps which had higher number of user reviews (i.e. \(\geq 500\)) were selected. For apps that had a detailed set of functionalities, user ratings and reviews were considered. Later, the products were categorized with Table \ref{tab:AppData} showing the categories, product names and a list of five similar apps for each product. For the data collection process, we scraped the data manually using a Github repository\footnote{facundoolano/google-play-scraper} (as the  google API is exclusive to app owners). Thus, we could only scrape at most 4,480 reviews per app.

\begin {table*}
\scriptsize
\centering
\caption{The list of included Apps in our study}
\vspace{-2mm}
\label{tab:AppData}
\begin{tabular} {p{3.95cm}|p{13.95cm}} \hline
{\bf Category [Name of App]} & {\bf List of Similar Apps on Google Play}\\
\hline

{Learning {\bf [Study Guide App]}} & (1) Quizlet learn with Flashcards - Quizlet LLC, (2) Study - White Sun apps, (3) Todait - Smart Study Planner - Todait Inc, (4) Student Agenda - Appsbuyout Development, (5) Time to Study - Sergio Arnillas\\
\hline
{Productivity {\bf [Workflow and Productivity App]}} & (1) Evernote - Stay organised- Evernote Corpopration, (2) Todoist To-Do List, task List-Doist, (3) Slack - Slack Technologies Inc, (4) Skype for Business for Android - Microsoft Corporation, (5) Trello - Trello Inc\\
 \hline
{Games {\bf [Guess Who App]}} & (1) Heads Up! - Warner Bros. International Enterprises, (2) Guess Who? Cards by Shuffle - Cartamundi Digital, (3) What am I? - 2minds Dev, (4) GuessUp - Party Charades, (5) CharadesApp - What am I? - artGS\\
 \hline
{Recommendations {\bf [Board Game Recommendation App]}} & (1) GameFindr - MTGHeadQuarters, (2) BoardGameGeek - BGG4Android, (3) Game rules - BANJEN Software, (4) Board Game Companion - Gadamagaska, (5) ScorePal - happyFaceDevs\\
 \hline
{Events {\bf [UofC Events App, Event Parser App ]}} & (1) Events near Me - Dimitris Konomis, (2) Eventbrite - Fun Local Events - Eventbrite, (3) Eventtus - Events App - Eventtus, (4) Meetup - Meetup, (5) StubHub - Events tickets - StubHub\\
\hline
{Service {\bf [Stumped]}} & (1) Tutor.com To Go - Tutor.co, (2) Chegg Tutors: Online Tutoring - Chegg, Inc, (3) TutorGPS Live Online Tutoring- Apps from Web, (4) Skooli Online Tutoring - Skooli Inc., (5) Varsity Tutors Live Tutoring - Varsity Tutors LLC\\
\hline
{Fitness {\bf [UofC Fitness App]}} & (1) Fitbit - Fitbit, Inc., (2) Google Fit - Fitness Tracking - Google Inc, (3) FitNotes - Gym Workout Log - James Gay, (4) Abs workout - Caynax, (5) Goal Tracker and Habit List - Intrasoft\\
 \hline
{Data Collection {\bf [Express Food Ordering App]}} & (1) Food Delivery by DoorDash - DoorDash, (2) UberEATS:Food Delivery - Uber Technologies, Inc, (3) SkipTheDishes - Food Delivery - Team SkipTheDishes, (4) Foodoa - Finest Food Delivery - Foodora GmbH, (5) Just Eat - Order Food Delivery - Just Eat Holding Limited\\
\hline

 \end{tabular}
 \vspace{-5mm}
\end{table*}

The data returned from Google App Review was is CSV format and was not immediately ready for analysis. To prepare the corpus for our algorithms we converted each row into a text file. We iteratively performed the following steps to clean and prepare the corpus for automatic data analysis.  
 
\begin{labelledsteps}
\item [\it Convert Alphabetical Text to Lowercase] We removed case sensitivity to ensure we don't analyze the capitalization of a word as a separate case than its lower-case counterpart.
\item [\it Removing numbers and punctuations] All rating numbers and punctuations generated after converting CSV to text files were removes at this step.
\item [\it Removing stopwords] Stopwords are common words that provide no meaning on their own, such as ``the''. Here we removed the default set of stopwords in the {\bf tm\_map} package of {\bf R}. Moreover, we added a new set of stop words specific to the context of mobile app analysis such as {\it application/app, download}, and {\it play}.
\item [\it Manual Transformation] We replaced words that were synonymous to each other since they would either appear too scattered due to their independence. Some of these words had to be united before the corpus was stemmed. For instance, we replaced `any timeÕ with `anytimeÕ.
\item [\it Strip Whitespace] We removed excessive whitespace such as newlines, double 
spaces and tabs.
 
\item [\it Stemming] Stemming is the process of reducing words to their origins. For example, ``easily'', ``easy'' and ``easeÕÕ' would all become ``ease'. 
\end{labelledsteps}

\subsection{Data Analysis}
The data analysis phase consisted of two sub-phases: first the data gathered from Play Store was analyzed by applying the Latent Dirichlet Allocation (LDA) algorithm to classify reviews based on the frequency of word co-occurrences. A topic, as defined in the LDA approach, is a probability distribution over a vocabulary \cite{LDA}, for which the {\bf topicmodels} package in {\bf R} was needed for the correct implementation of the LDA algorithm. This procedure also required the Gibbs sampling option, as it provides greater accuracy than the variational algorithm \cite{Gibs}. Moreover, to apply the LDA approach, the number of topics ({\it k}) which will be used by the algorithm for classifying the reviews should be defined upfront. For each category of apps, we ran the algorithm for different values of {\it k} (i.e. {\it k=2..7}) and made a choice as to the most suitable {\it k-value} based on our evaluation of the results. As illustrated in Table \ref{tab:Results}, the value of {\it k} differs between app types. 

Next a manual data analysis was performed on a random sample from each app category. 500 reviews (\(4,000\) in totall) were randomly selected from each category, followed by a manual coding process using the web-based coding tool Saturate (Section \ref{sec:RQ1-Analysis}). The detailed results of each step in this phase are found in Section \ref{sec:RQ2-Results} and Table \ref{tab:Results}.

\subsection {Results}

Table \ref{tab:Results} presents the results of the automatic and manual data analysis performed on the app review datasets. The second column of the table presents the number of explored topics (i.e. {\it k}) for each category. For instance, the results of applying the topic modelling approach on FlashCard apps revealed two topics: (1) Satisfaction (use, great, help, love, work) and (2) Functional Requirements (e.g. make, deck, like, time, set). Our first clue in determining this category is the frequency of the word ÔmakeÕ in this topic followed by what, in the context of a flashcard app comment dataset, seemed like functions and features of the app. The Satisfaction topic the most prominent and the one that remained constant as the first topic for {\it k} values as high as 5. In essence this category represents self-reported instances of FR acceptance by the app clients. 

The results of the manual data analysis for the same app category (i.e. Flash Card) revealed an additional category; that of {\it bug reports}. While this was a smaller category, the manual coding done with the Saturate app did reveal it to be significant. Items in this category mainly concerned bugs that would cause users to loose all of their flashcards. The presence of the word ÔlikeÕ in topic 2 was explained during the manual data analysis by the occurrence of many reviews that started out praising the app before making some suggestion about an additional feature.
The manual analysis also revealed the nature of captured FRs. One example of this is the large number of requests to be able to see the answer first and guess the question. 
 
For the ``Games'' category, topic modelling revealed two categories. The first of these appeared to be for FRs but upon inspection of the document to topic assignment of the LDA algorithm, was determined to to represent {\it bug reports}. The second topic determined by LDA was Satisfaction, as defined above. Manual analysis of a random sampling of the data set concurred with these two primary topics, but it should be noted that there were a small number of FR requests. Comments such as Ôjust work on controlsÕ are representative of this.

\label{sec:RQ2-Results}
\begin{table*}
\vspace{-2mm}
\centering
\scriptsize
\caption{Results of the LDA and Manual Data Analysis [k represents the number of topics resulted from applying the LDA approach] }
\vspace{-3mm}
\centering
\rowcolors{1}{ }{Gray}
\begin{tabular}{p{3 cm}p{0.5 cm}p{3.5cm} p{4.1cm}p{6.2cm}} 
\hline
{ \bf  Category [\# of Reviews*]}&{\bf k} & {\bf User review's themes (LDA)}&{\bf Manual Data Analysis [500 reviews] }&{\bf Sample Requirements}\\ \hline
{\bf Learning Apps [\(8,477\)]}&2&Satisfaction, FR&Satisfaction, FR, Bug Reports&Search feature,  Jeopardy question style option, Add directory structure\\
{\bf Games [\(5,670\)]}&2&Satisfaction, FR&Satisfaction, Bug Reports&Spanish language support, More levels/words, Louder game sounds\\
{\bf Study Guide [\(5,651\)]}&3&Usability, Satisfaction, FR&Usability, Extensibility, Satisfaction&Addsearch function, move cards across  studying sets\\ 
{\bf Service [\(199\)]}&2&Relibaility, Usability&Reliability, Usability, Extensibility& Ensure similar features across platforms, create easy to use UI  \\ 
{\bf Recommendation [\(998\)]}&3&Operability, Satisfaction&Reliability, Usability&Browsing Game Collection, User-friendly Interface, Random Selection\\    
{\bf Events: [\(13,189\)]}&2&Usability, Reliability&Useability, Searchability&Removal of ads, Irrelevant recommendation, Having access to functionality that is exclusive to website\\    
{\bf Productivity [\(22,335\)]}&3&Usability, FR, Satisfaction&Usability, Operability&Call back functionality,sound quality,Add colour note feature\\
{\bf Fitness: [\(19,743\)]}&3&Availability, Reliability, Usability&Reliability, Privacy, Resource Usage&Better bluetooth connectivity with external devices, Information and recommendation on nutritional food\\                                                                                                                                                         
\hline    
\rowcolor{white}
{\it * Total = \(70, 592\) reviews }&&&&\\                                               
\end{tabular}
\vspace{-6mm}
\label{tab:Results}
\end{table*}

\subsection{Lessons Learned and Challenges} 
 
The results of this experiment show that while user reviews are indeed a powerful tool for capturing FRs, this is not without cost; exhibited in the prevalence of bug reports in certain app categories. That WOz was as effective as it was in capturing NFRs and clarifying existing FRs, combined with its low cost, means that it would be a mistake to dismiss this method. That it is effective has been empirically shown in this study. Moreover, considering the lack of visualization techniques for modelling NFRs \cite{Viz}, the WOz technique can be used as a supplementary approach to visualize a number of NFRs (e.g. usability and learnability).  
Additionally, user reviews often do not express requirements unambiguously and provide little opportunity for clarification. The converse is true for WOz, as it allows for the instant feedback of client acceptance of some requirements and clarification of others.

However, WOz is not without its faults. Specifically, NFRs like Privacy, Reliability, Operability, Availability, and Resource Usage cannot be captured by WOz technique. In comparison to user reviews, WOz is slow to adapt to changes in requirements; itÕs strength lying primarily in the early stages of development and the clarification of requirements.

This study indicates that there is value in both of the requirement capturing techniques. As is often the case, the challenge lies in finding the golden middle way; an optimal balance between expedience and quick reaction to market conditions (reactive) vs. getting the requirements right the first time (proactive).

\label{sec:LL2}
\subsection{Threats to Validity}
Conducting both studies on particular app domains might question the generalizability of our findings. We tried to address this threat by analyzing a fairly large dataset (i.e. 13 app development teams, 40 Android apps on Google Play, and \(70,592\) reviews). This helped us triangulate findings obtained through the WOz approach with the results from the follow-up automatic and manual app review analysis method. Moreover, our data preparation poses another threat to the validity of our results. Namely, defining context-based stop words can impact the results of topic modelling. We attempted to mitigate this threat by iteratively revising the list of stop-words. 

A recent study by Shakeri H. A. et al. \cite{Zahra2} shows that LDA performs poor for extracting topics from short requirements documents, as the one typical for mobile app reviews. This might question the validity of the results of RQ2. However, since we were only looking for the main themes of user reviews and were not concerned about the details of each class, this limitation of the LDA approach does not significantly impact our results.  Moreover,  we manually analyzed {\it 4k} of app reviews (i.e. 500 reviews for each app category) to validated the results of the LDA approach and to minimize the impact of our data analysis errors on the main findings of our study.



\section{Conclusion and Future Work}
To study the application of the WOz technique in capturing mobile app requirements and to compare the ability of this method with user review analysis, we conducted two studies: a field study on 13 mobile app development teams and a retrospective data analysis on mobile app reviews (\(70,592\) reviews of 40 apps available on Google Play (Table \ref{tab:AppData}. From both studies we found that while each of WOz and app review analysis techniques can be applied to capture specific types of requirements, an integrated requirements elicitation process including both methods would reduce the communication gap between users and developers in early stages of the development process and mitigate the risk of requirement changes in later stages.
 
In future work we will aim to extend our investigation of the WOz technique by looking at other domains of mobile applications and to replicate our automatic topic modelling by applying modelling techniques appropriate to short text documents (e.g. Biterm Topic Modeling (BTM)).  Moreover, analyzing user reviews of applications which deployed the WOz technique at early stages of their development process (e.g. as in our ongoing app development project \cite{RETTA}), is another goal for replicating this case study. 
\vspace{-1mm}

\bibliographystyle{IEEEtran}
\bibliography{IEEEabrv,REW'17}
\end{document}